# Resonances and Quantum Scattering for the Morse Potential as a Barrier.

G. Rawitscher, Cory Merow, Matthew Nguyen, and Ionel Simbotin.
Department of Physics, University of Connecticut,
Storrs, CT 06269-3046

November 20, 2001

## Abstract

Quantum scattering in the presence of a potential valley followed by a barrier is examined for the case of a Morse potential, for which exact analytic solutions to the Schrödinger equation are known in terms of confluent hypergeometric functions. For our application the potential is characterized by three parameters: the height of the barrier, the distance of the barrier from the origin of the radial variable $r$, and a diffuseness parameter. The wave function, defined in the interval $0 \leq r < \infty$, is required to vanish at $r = 0$, and hence represents a radial partial wave for zero angular momentum. The vanishing at $r = 0$ requires a special combination of hypergeometric functions, and can lead to resonances for incident energies which occur below the top of the barrier. Numerical values for the analytical phase shifts are presented in and outside the resonant regions, and the corresponding properties of the scattering S-matrix are examined in the complex momentum plane, mainly for pedagogical reasons. The validity of the Breit-Wigner approximation to the resonant phase shifts is tested, and the motion of a "resonant" wave packet slowly leaking out of the valley region is also displayed.

## 1  Introduction.

The Morse potential, introduced in the 1930's [1], is one of a class of potentials [2] for which analytic solutions of the Schrödinger equation are known. The Morse potential is widely used in the literature as a model for bound states, such as the vibrational states of molecules. For these applications the potential is defined as a function of the variable $x$ which ranges from $+$ to $-\infty$, and is given in terms of two parameters $V_0$ and $\alpha$ by

$$V_M(x) = V_0 e^{-x\alpha}\left(2 - e^{-x\alpha}\right) \quad (1)$$

where $V_0$ is negative. At $x = 0$ it has a negative (attractive) minimum of depth $V_0$, and it goes smoothly to zero in the limit of large $x$. For $x$ less than $-\ln(2)/\alpha$ this potential becomes positive (repulsive), and as $x$ decreases even further, the potential becomes increasingly large. This potential will be denoted as the "regular" potential in what follows. In the present application, by contrast, the sign of the overall potential is reversed ($V_0 > 0$) so that the attractive valley turns into a repulsive barrier, giving rise to the "inverted" potential. In addition the variable $x$ is replaced by the radial variable $r = x + r_e$, defined only in the interval $0 \leq r < \infty$,

$$V_M(r) = V_0 e^{-(r-r_e)\alpha} \left(2 - e^{-(r-r_e)\alpha}\right), \qquad (2)$$

where $r_e$ is the location of the top of the barrier. This three-parameter form the Morse potential has been given before [3], and an example is illustrated in Fig. 1. This case, where there is a valley followed by a barrier, will be denoted as the "inverted" Morse potential.

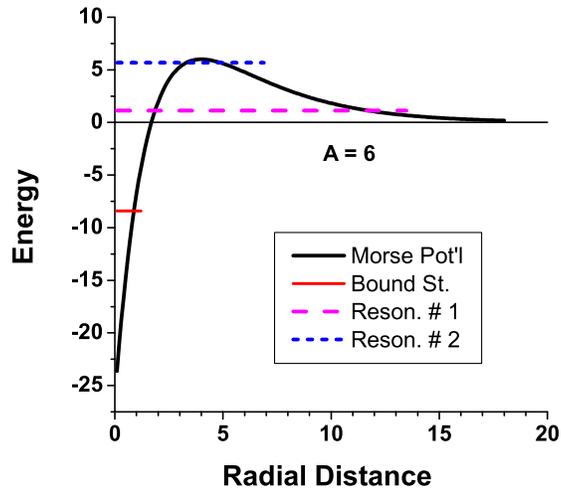

Figure 1: The Morse potential for the case $A = 6$ is illustrated by the thick line. The other parameters are $r_e = 4$, and $\alpha = 0.3$. The bound state and the two resonances are indicated by the horizontal lines. The energy is in units of inverse distance squared ($fm^{-2}$), and the radial distance is in units of $fm$.

Much work has been done with the "regular" Morse potential: on algebraic methods to either generate bound eigenstates [4] or scattering states [5]; on the analytic connection between the states in the regular and the inverted Morse potentials [6]; on approximations to barrier penetration [7], [8]; and on shape resonances [9]. The analytic solutions of the Schrödinger equation for the Morse potential can also be found in several textbooks [10], but not in the same detail as in the references above. Most of these investigations limit themselves to examining the analytical properties of bound states, and of the respective ladder operators which generate one state from the previous one, but do not illustrate the results by means of numerical values.

Our study differs from the ones above, in that we use the three-parameter "inverted" form of the potential and impose the condition that the wave function vanish at $r = 0$. This valley-barrier combination, together with the reflection of the wave function at the origin, produces conditions which can generate resonances. We numerically illustrate the properties of the resonances by evaluating the analytical results for the phase shifts and the wave functions for real momenta, and the S-Matrix in the complex momentum space. Our Morse potential is much closer than a square well to applications which have physical content, such as the interactions between nuclei at low (astrophysical) energies, or between atoms where barriers and resonances are involved. The resonances will be described both in terms of the rapid variation of the phase shift as a function of the incident (real) momentum $k$, and in terms of the poles of the scattering $S-matrix$ in the complex momentum space $k = k_R + i\, k_I$. One of the purposes of the present study is to provide a pedagogical illustration of the properties of resonances by means of the analytic solutions of the Schrödinger equation with a reasonably realistic potential. Another aim is to provide precise values for the phase shifts in the vicinity of a resonance, based on the exact analytical results, to be used for comparison values in tests of numerical methods to solve the Schrödinger equation [11]. The resonance region provides a specially stringent numerical test case, because at resonance the wave function in the interior of the valley forms a standing wave of large amplitude which then *decreases* through the barrier region, a feature which is difficult to reproduce numerically with great accuracy. This is because small errors in the numerical evaluation of Eq. (3) are amplified since they increase exponentially as they propagate through the barrier region, in contrast to the true solution which decreases exponentially. These errors become larger the more narrow the resonance, since the decrease of the true wave function then becomes more pronounced. These errors become larger the more narrow the resonance, since the decrease of the wave function then becomes more pronounced, and they do not manifest themselves outside of the resonance region, since there the true wave function also increases.

The radial, one-dimensional, Schrödinger equation to be solved is

$$\left(\frac{d^2}{dr^2} + k^2\right) R(r) = V(r)\, R(r), \tag{3}$$

where $k$ is the wave number of the incident particle, related to the incident energy $\overline{E}$ according to

$$k^2 = (2m/\hbar^2)\, \overline{E} = E, \tag{4}$$

and $V$ is the Morse potential $V_M$ re-scaled by the mass factor $2m/\hbar^2$, i.e.,

$$V(r) = A\, e^{-(r-r_e)\alpha} \left(2 - e^{-(r-r_e)\alpha}\right). \tag{5}$$

The three parameters are thus

$$A = 2m/\hbar^2 \, V_0, \ r_e \text{ and } \alpha. \tag{6}$$

The units of $A$ is inverse length squared, $r_e$ is in units of length, and $\alpha$ is in units of inverse length. The radial wave function $R$ is required to vanish at the origin and is normalized such that asymptotically it equals

$$R(r \to \infty) = \sin(kr + \delta), \tag{7}$$

where $\delta$ is the phase shift.

In section 2 we derive the analytic results, in section 3 we focus on the properties of the resonances, in particular we study the extent of the validity of the Breit-Wigner approximation to both the poles in the complex momentum space, as well as to the k-dependence of the phase shift in real momentum space. Section 4 describes the construction of a "resonant" wave packet in order to illustrate the time delay of the packet in the valley region, and Section 5 contains the summary and conclusions.

## 2 Analytic Results.

As is well known [1], [7], solutions for Eq. (3) can be expressed in terms of confluent hypergeometric functions $M(a, b; z)$ of the type $_1F_1$, defined in Eq. (13.1.2) of Ref. [12], where $z$ is a purely imaginary variable defined as

$$z = z_0 \, \exp(-r\alpha), \ with \ z_0 = 2\,i\,\beta\,\exp(r_e \alpha), \tag{8}$$

and where

$$\beta = (A)^{1/2}/\alpha. \tag{9}$$

The functions $M$ can be defined as usually convergent series in powers of the variable $z$, with coefficients which depend on the parameters $a$ and $b$. The solution $R(r)$, which vanishes at the origin and which obeys the normalization condition (7) is given by an appropriate combination of the functions $M$ according to Eq. (11), as is shown in Appendix 1. As is also shown there, this combination can be written most conveniently in terms of the function $\mathfrak{M}(k,z)$, defined as

$$\mathfrak{M}(k,z) = e^{-z/2} M(1/2 + ik/\alpha - i\beta, 1 + 2ik/\alpha; z), \tag{10}$$

with the result

$$R(r) = \frac{1}{2i} |\mathfrak{M}(k, z_0)| \, \left\{ [\mathfrak{M}(k,z)/\mathfrak{M}(k,z_0)]^* \, e^{ikr} - [\mathfrak{M}(k,z)/\mathfrak{M}(k,z_0)] \, e^{-ikr} \right\}. \tag{11}$$

The phase shift $\delta$ is the phase of $\mathfrak{M}(z_0)$, i.e.,

$$\exp[i\delta(k)] = \mathfrak{M}(k, z_0)/|\mathfrak{M}(k, z_0)|, \tag{12}$$

the symbol ∗ denoting complex conjugation. For real values of $k$ both the function $R$ and the phase shift $\delta$ are real, due to the validity of Eq. (42) derived in Appendix 1. These results illustrate the important rôle played by the point $r = 0$. This point, represented by $z_0$, is absent from the expressions for the bound-state eigenvalues and eigenfunctions of the usual (non-inverted "regular") Morse potential (1), since these expressions are not subject to the condition that the wave function vanish at $x = -r_e$. Equations (11) and (12) have not been given before. It is remarkable that the same analytical expression, Eq. (11), can at the same time represent the oscillatory nature of the function $R$ in the valley region, the exponential nature of the function in the barrier region, and finally give rise to the correct asymptotic sinusoidal behavior for large values of $r$.

For complex values of $k$ the phase shift becomes complex, and the absolute value of the scattering $S-matrix$ is no longer unity but can have poles in certain locations in the complex $k-$plane. A convenient expression for the $S-matrix$ is then given by

$$S(k) \equiv e^{2i\delta} = \frac{M_0^{(+)}}{M_0^{(-)}}, \tag{13}$$

where

$$M_0^{(\pm)} = M(1/2 \pm ik/\alpha - i\beta, 1 \pm 2ik/\alpha; z_0).$$

For complex values of $k$ Eqs. (11) and (12) are no longer valid, and will be replaced by equations given in Appendix 1.

It is well known that the $S-matrix$ satisfies the two important properties

$$[S(k^*)]^* \, S(k) = 1 \tag{14}$$

and

$$S(k) \, S(-k) = 1. \tag{15}$$

It is shown in Appendix 1 that our Morse $S-matrix$, Eq. (13) satisfies both properties. The first equation, which implies that $|S(k^*)| \times |S(k)| = 1$, shows that if $S$ has a pole at the point $k_0 = k_r + i\,k_i$, then $S$ has a zero at the "mirror point" about the real k-axis, $k_r - i\,k_i$. The combination of both equations implies that $S(-k^*) = [S(k)]^*$, which shows that if $S$ has a pole at $k_0 = k_r + i\,k_i$, then it also has a pole at the mirror point about the imaginary axis at $(-k^*) = -k_r + i\,k_i$. The latter property was clearly illustrated in a seminal paper by Nussenzweig [13], who examined the poles of the S-matrix for a square well potential.

## 3 Resonances

Resonances play an important role in many branches of physics. The book by Kukulin, Krasnopol'sky and Horáček [14] gives an extensive description of the theories used to calculate quantum resonances, mainly for coupled systems of equations. Single channel resonances for a potential with an attractive valley and a repulsive barrier of the type described here, occur for certain narrow regions of the (real) incident positive energy above zero, but still below the top of the barrier. The corresponding wave function at the interface between the valley and the onset of the barrier region (i.e., the inner turning point) has a logarithmic derivative such that in the barrier region the wave function has to decrease. These resonances are thus extensions of the bound states, for which the wave function also has to decrease in the classically forbidden region.

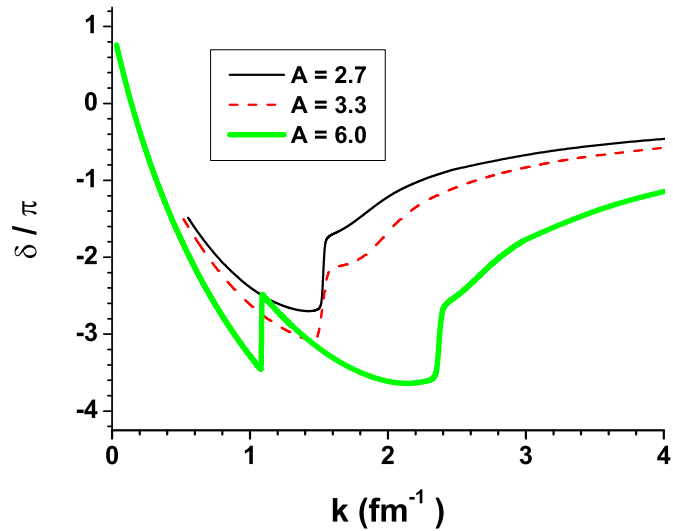

Figure 2: The phase shift in units of $\pi$, as a function of the incident momentum $k$, for three values of the potential strengths $A$. The units of $A$ and $k$ are $fm^{-2}$ and $fm^{-1}$, respectively.

The Morse potential for $A = 6\,fm^{-2}$, $r_e = 4\,fm$, and $\alpha = 0.3\,fm^{-1}$, illustrated in Fig. 1, has one bound state at $E_B = -8.1090$, and two resonance energies at $E_1 = 1.1783$, and $E_2 = 5.626\,fm^{-2}$. The latter is already very close to the top of the barrier, $6\,fm^{-2}$. All the results illustrated in the figures were obtained by numerically evaluating the analytic expressions given in Section 2, both by using MATHEMATICA [15] and by constructing a FORTRAN program for checking purposes. The key ingredient in the latter was the evaluation of the Hypergeometric Function $M$ by means of its power series expansion, using quadruple precision. Care was taken that the cancellations between powers did not introduce errors beyond the desired prescribed precision. The results for real values of $k$ were also checked against a numerical method of solving the Schrödinger equation [11]

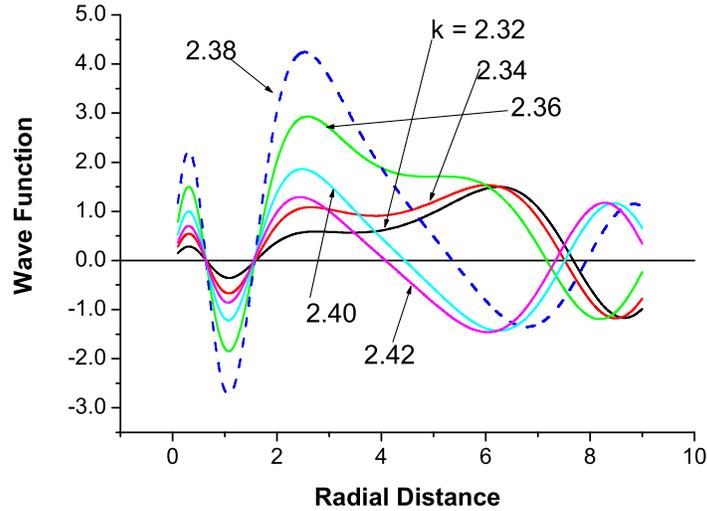

Figure 3: Wave functions for various values of $k$ in the region of the second resonance for the $A = 6$ Morse potential. As $k$ increases, the wave functions are "pulled" towards the origin, as can be seen by observing how a node moves through the barrier to the left.

One of the most striking features of a resonance is the rapid variation of the phase shift with the incident momentum, as it sweeps over the resonance. This behavior is illustrated in Fig. 2, for various values of the potential strength $A$. As the strength is changed, the location of the resonances changes accordingly. In the region of the resonances the phase shifts *increase* with $k$. This behavior is intimately related to the time delay between the arrival of an incident wave packet and the re-emergence of the resonant part of the wave packet from the interior of the potential. That time is related [16] inversely to the width $\Gamma$ of the resonance, which in turn is related to the distance of the pole of the $S-matrix$ from the real axis, as is well known and as will be illustrated further below.

Another important feature of a resonance is that the magnitude of the wave function in the valley region becomes large for momenta in the vicinity of the resonance. This is illustrated in Fig. 3 for the second resonance of the $A = 6$ case, which occurs in the vicinity of the momentum $k \simeq \sqrt{5.626} = 2.37\, fm^{-1}$. The wave function which has the largest amplitude near the inner turning point occurs for $k = 2.38$, all the other wave functions, either below or above this $k-$value, having smaller amplitudes.

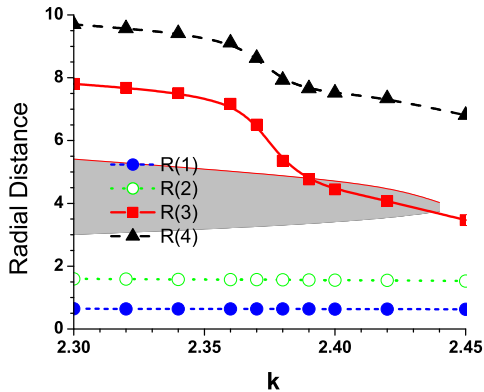

Figure 4: Radial positions of the first four nodes of the wave functions in the resonance region, in units of $fm$, are shown on the vertical axis. The corresponding momentum values $k$, in units of $fm^{-1}$, are displayed on the horizontal axis. The shaded area represents the barrier region. It is delimited by the outer and inner turning points of the barrier. Points which fall within this area correspond to zeros located in the barrier region, which is the case for the curve $R(3)$ of the third node. The value of $k$ where $d[R(3)]/dk$ is largest, approximately at $k \simeq 2.38$, corresponds to the center of the resonance.

A closely related feature is the position of the zeros of the wave function. As the value of $k$ increases from below the resonance ($2.32\,fm^{-1}$) to above ($2.42\,fm^{-1}$), the first minimum of the wave function beyond the barrier region moves to the left across the barrier, until, when $k$ is above the resonance, another half wavelength is added to the wave function in the valley region. This can also seen by observing the trajectory of the zeros of the wave function displayed in Fig. 3, and is illustrated more clearly in Fig. 4. The shaded region displays the location of the barrier, its upper and lower borders being the locus of the outer and inner turning points, respectively. Curve R(3), for the third zero, is the most interesting one. It shows that the zero of the wave function closest to the right of the outer turning point rapidly sweeps to the left through the barrier region into the valley region, a feature which will be made use of for a new method to find the location of a resonance.

The simplest method of finding a resonance consists in calculating the wave function and observing the number of nodes inside of the valley region. As discussed above, that number changes from one resonance to the next, just like it would for the case of bound states. However, for narrow resonances this method is quite cumbersome. A speedier numerical method is based on the second observation made above, i. e., only near a resonance can a zero of the wave function occur in the barrier region. One can make use of this observation by numerically finding the eigenvalues $k^2$ of the equation $\left(-\frac{d^2}{dr^2} + V(r)\right) R(r) = k^2 R(r)$, constrained by the two-point boundary condition that the wave function vanish both at the origin and at a point $r_B$ somewhere in the vicinity of the barrier region. By varying $r_B$ and obtaining the corresponding eigenvalue $k$ one can trace out curves of the type displaced in Fig. 4. For example if $r_B = 4\,fm$, then for the $A = 6$ case one finds [17] two resonances, which correspond to the values of $k$ of 1.085 and 2.426. The latter is in excellent agreement with the value of $k = 2.425$ obtained from Fig. 4, which is the point where the $R(3)$ curve intercepts the $r = 4$ line. The first resonance is very narrow, and is discussed further below in connection to the poles of the S-matrix. This method of finding the resonances has the advantage that only the eigenvalues need to be known, and not the wave function or the asymptotic phase shifts. A more detailed account of this method will be described in a future study.

There are several other methods to locate resonances. A very general one, especially useful for the case of double or triple peaked barriers, makes use of the spectral energy density method [18], and was recently applied to the resonance levels of atoms in an external electric field [19]. An even more general one makes use of the set of Siegert functions in order to calculate the position of the poles of the $S-matrix$ in complex momentum space, by establishing and then solving a certain eigenvalue equation [20].

## 3.1 Poles of the $S-matrix$, and the Breit-Wigner approximation.

In the present study the behavior and location of the S-matrix poles in complex momentum space for the Morse potential case are investigated numerically, by evaluating the analytic expression, Eq. (13) repeatedly, until the location of the pole is found to within the desired accuracy. The poles for the first and second resonances in the $A = 6$ case were thus found at

$$k_R^{(1)} = 1.507160786\,fm^{-1}, \ \ k_I^{(1)} = -3.440531 \times 10^{-6}\,fm^{-1}, \qquad (16)$$

and

$$k_R^{(2)} = 2.371752656\,fm^{-1}, \ \ k_I^{(2)} = -0.007406315\,fm^{-1} \qquad (17)$$

respectively. Other, more sophisticated, methods could also be considered, such as locating the zeros of the denominator in Eq.(13) by an iterative Newton-type method, or by some other means [19], [20]. In order to demonstrate that the positions of a pole of the S-matrix and that of a zero are located at points mirrored across the real axis in k-space, the contour lines of the logarithm of the absolute value of the S-matrix for the resonance #2 (closest to the top of the barrier) for the $A = 6$ case are shown in Fig. 5. A three dimensional view of that pole is shown in Fig. 6. A rich structure of poles is also present at energies above the top of the barrier, as is illustrated in 7. These poles contribute collectively to the non-resonant value of $\delta(k)$, for $k \geq 2.50$. Similar non-resonant poles are responsible for the shoulders which occur to the right of the resonances for the cases $A = 2.7$ and $3.3$, displayed in Fig. 2. As $A$ increases from $3.3$ to $6.0$, the energy of one of these non-resonant poles becomes submerged below the top of the barrier, giving rise to a new resonance.

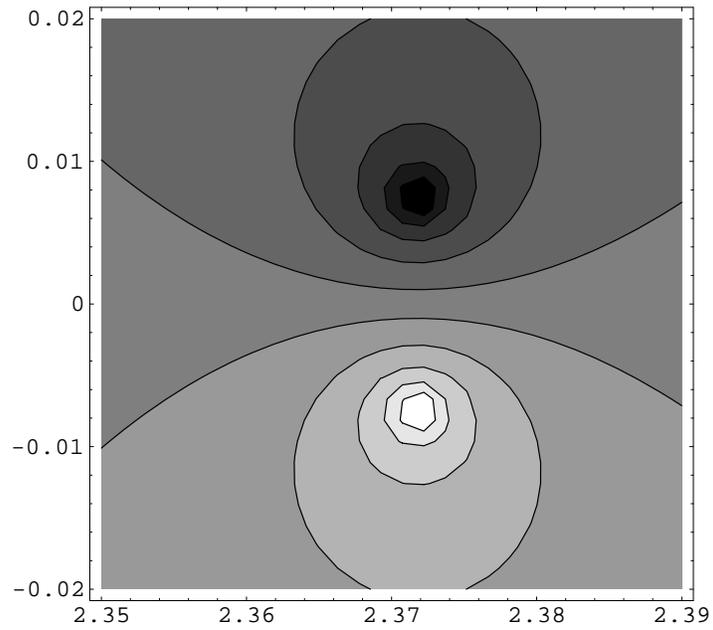

Figure 5: Contour plot of the pole and adjoining zero of the absolute value of the $S-matrix$ in momentum space of the second resonance for the $A = 6$ case. The horizontal (vertical) axis contains the real (imaginary) part of the momentum $k$, in units of $fm^{-1}$. The contour lines are spaced in equal increments of the logarithm of $|S|$, so as to display the zero more clearly. The latter lies in the upper half of the plane.

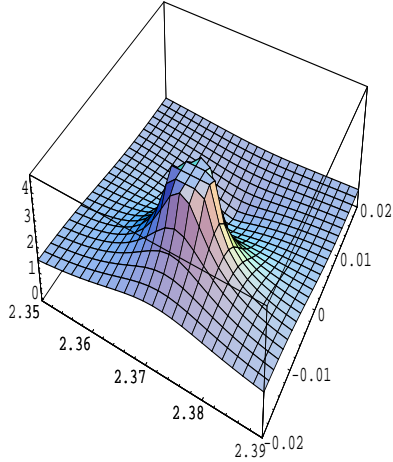

Figure 6: The absolute value of the $S-matrix$ in the complex momentum plane for the second resonance of the $A = 6$ case. The horizontal axis which is nearly parallel (orthogonal) to the plane of the paper contains the real (imaginary) part of $k$. The corresponding contour lines are displayed in Fig. 5.

Next the validity of the Breit Wigner approximation [21] will be discussed. In this approximation, the phase shift $\delta$ is decomposed into a non-resonant "background" part $\xi$ and a resonant part $\Delta$

$$\delta(k) = \xi(k) + \Delta(k). \tag{18}$$

The value of $\Delta$ changes rapidly with the incident energy $E$ according to the Breit-Wigner expression

$$\Delta_{BW} \simeq \arctan \frac{\Gamma/2}{E_r - E}. \tag{19}$$

In the above, $E_r$, and $\Gamma$ are the resonant energy and the width of the resonance, respectively, and $E$ is the incident energy, related to the incident momentum $k$ according to Eq. (4). In the above equation, all three quantities are in the same units of inverse length squared. The quality of this approximation for the second resonance for the $A = 6$ case is illustrated in Fig. 8, B the solid and dashed lines represent the exact and the Breit-Wigner results, respectively. The value of $E_r$ and $\Gamma$ are not adjusted, but are obtained from the

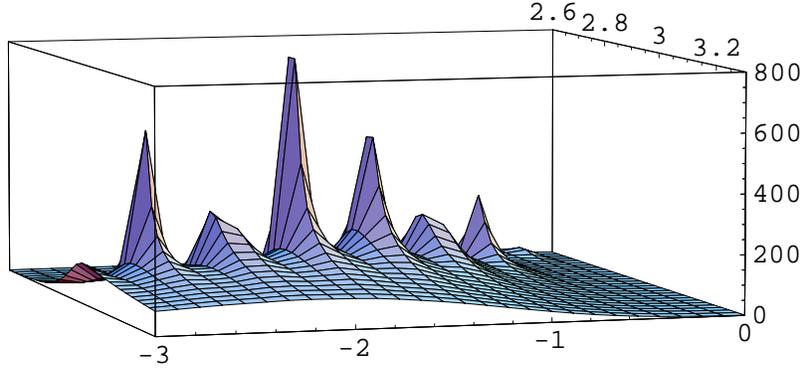

Figure 7: The absolute value of the $S-matrix$ in the part of the momentum plane above of the top of the barrier, at $A = 6$ ($\sqrt{6} = 2.45$). The horizontal axis which is nearly parallel (orthogonal) to the plane of the paper contains the imaginary (real) part of $k$.

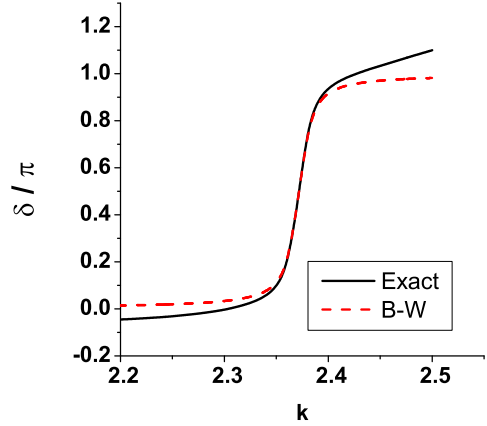

Figure 8: Accuracy of the Breit-Wigner approximation to the second resonance for the $A = 6$ case. The analytic result (solid line) was adjusted to agree with the B-W approximation at the central point $k_r = \sqrt{E_r}$. The width $\Gamma$ and the position $E_r$ in the B-W expression Eq. (19) are not adjusted, but are taken from the location of the pole of the $S-matrix$, according to Eqs. (23) and (17).

values of $k_R^{(2)}$ and $k_I^{(2)}$ given above (17) and making use of Eq. (23) given below. The value of the non-resonant part $\xi$ of is adjusted by bringing the analytic and the Breit-Wigner results into agreement at the center of the resonance, at $E = E_r$ where $\Delta = \pi/2$. The agreement near the center of the resonance is very good. The pronounced disagreement for $k > 2.4\,fm^{-1}$ is due to the proximity of the poles which lie above the barrier, but which nevertheless are quite close to the top of the barrier, as is shown in Fig.7

The Breit-Wigner approximation (19) can be extended into the complex momentum $k-$plane near the location of the pole, as follows. If one makes use of the identities

$$e^{i\Delta}\sin(\Delta) = \frac{1}{2\,i}\left(e^{2\,i\,\Delta} - 1\right) = \frac{\tan(\Delta)}{1 - i\,\tan(\Delta)}$$

together with Eq. (19), valid for real momenta, one obtains

$$\frac{1}{2\,i}\left(e^{2\,i\,\Delta} - 1\right) \simeq \frac{\Gamma/2}{E_r - E - i\,\Gamma/2} \equiv \frac{\Gamma/2}{E_P - E}, \qquad (20)$$

according to which the Breit-Wigner approximation to the resonant part of the $S-matrix$ for real momenta becomes

$$e^{2i\Delta} \simeq \frac{E_r - E + i\,\Gamma/2}{E_r - E - i\,\Gamma/2} = \frac{E_P - E + i\,\Gamma}{E_P - E}. \qquad (21)$$

In the above, $E_r$, and $\Gamma$ are real constants, and $E_P$ is the complex quantity

$$E_P = E_r - i\,\Gamma/2. \qquad (22)$$

When $E$ is real, then it follows from Eq. (21) that $|\exp(2\,i\,\Delta)| = 1$, and hence $\Delta$ is also real. The analytic continuation of Eq. (21) into the complex $E-$ plane shows that $E_P$ is the position of a pole of the $S-matrix$, and Eq. (21) becomes Breit-Wigner's approximation to the $S-matrix$ for complex energies $E$ in the vicinity of the pole. From the location of the pole one can find the value of the real resonant energy $E_r$ by expressing the complex energy in terms of the square of the complex momentum. If in the momentum plane the pole occurs at $k_R^{(P)} + i\,k_I^{(P)}$, then using $E_P = \left[k_R^{(P)} + i\,k_I^{(P)}\right]^2$ together with Eq. (22) one finds

$$E_r = \left[k_R^{(P)}\right]^2 - \left[k_I^{(P)}\right]^2 \qquad (23)$$

$$\Gamma/2 = -2\,k_R^{(P)}\,k_I^{(P)}. \qquad (24)$$

The Breit-Wigner approximation to the $S-matrix$ near the pole, Eq. (21), is consistent with the discussion made after Eq. (14), according to which in complex momentum space a zero of the $S-matrix$ occurs at a mirror point above the pole. This can be seen as follows. According to Eq. (21) a zero of the $S-matrix$ occurs at the point $E_0 = E_r + i\,\Gamma/2$, and, according to Eq. (23) the corresponding values of the momenta are $k_R^{(0)} = k_R^{(P)}$, and $k_I^{(0)} = -k_I^{(P)}$.

However, according to the Eqs. (21) and (23), a pole would also occur at $k_R^{(P)\prime} = -k_R^{(P)}$, and $k_I^{(P)\prime} = -k_I^{(P)}$, since then the values of $E_r$ and $\Gamma/2$ remain unchanged. This is not true since it would violate identity (15), and hence the question arises for how large a region in the complex momentum plane around the pole is the Breit-Wigner approximation valid. This question is examined in Appendix 2. It was found that the B-W approximation is reliable for circular regions around the pole out to radii as large as twice the distance $k_I^{(P)}$ of the pole to the real $k-$ axis.

## 4  Wave Packet

In this section the "resonant" portion of a wave packet will be constructed, in order to further demonstrate the properties of a resonance.

For the present section, the solutions $R(r)$, defined in Eqs. (3) and (11), will be denoted by $\Phi(k,r)$ so as to bring out the dependence on the momentum $k$. A general wave packet is obtained by the superposition of the positive energy eigenfunctions for all momenta $k$, from 0 to $\infty$

$$\Psi(r,t) = \int_0^\infty C(k)\,\Phi(k,r)\,\exp(-ik^2 t)\,dk. \qquad (25)$$

In the above, $C(k)$ are the coefficients to be determined as described below, the energy $k^2$ is given in the units of $fm^{-2}$, as defined in Eq. (4), and the time $t$ is in units of $fm^2$, obtained by multiplying the time in units of seconds by $\hbar/2m$. This time unit will be denoted as $T$

$$T = 1\ fm^2. \qquad (26)$$

¿From the given value $\Psi(r,0)$ of the wave packet at $t=0$ the coefficients $C(k)$ are obtained according to

$$C(k) = \frac{2}{\pi} \int_0^\infty \Phi(k,r)\,\Psi(r,0)\,dr. \qquad (27)$$

This result follows from the Dirac delta function orthogonality property of the functions $\Phi$

$$\int_0^\infty \Phi(k,r)\,\Phi(k',r)\,dr = \frac{\pi}{2}\left[\delta(k-k') - \delta(k+k')\right].$$

In the present application the wave-packet will be chosen to be a Gaussian function, which at $t=0$ is given by

$$\Psi(r,0) = e^{-(r-R)^2/b^2}. \qquad (28)$$

Graphs of the corresponding wave packet showed that the peak of the packet travels a distance of about $5 fm$ in the time $10\,T$.

The resonant part of the wave packet of interest for the present discussion is the contribution to the integral in Eq. (25) from the momenta in the resonant region $[k_R - \Delta, k_R + \Delta]$

$$\Psi_R(r,t) = \int_{k_R-\Delta}^{k_R+\Delta} C(k)\,\Phi(k,r)\,\exp(-ik^2 t)\,dk. \tag{29}$$

The purpose is to demonstrate that the time dependence of this packet is indeed related to the time in which the particle is trapped in the potential well, which in turn is inversely related to the width $\Gamma$ of the resonance. According to Ref. [14], page 28, the time delay of a wave packet due to a resonance is $\tau_D = 4/\Gamma$, while the lifetime of the resonance, as obtained from Heisenberg's uncertainty principle, is $\Delta t = 1/(2\Gamma)$. The latter will be used as a unit of time

$$\tau = 1/(2\Gamma)\,fm^2 \tag{30a}$$

for the present investigation. In order translate these time units into more physically meaningful quantities, we will assume that the particle has the mass of a proton, $\simeq 938 MeV$. Then $T = 1\,fm^2$ corresponds to the time needed for the particle to traverse a distance of $9.5 fm$ with the speed of light, and a width $\Gamma$ of $1.9 \times 10^{-5}\,fm^{-2}$ corresponds to an energy of $3.9 \times 10^{-4}\,MeV$.

The width $b$ of the packet of Eq. (28) is chosen as $b = 2\,fm$, the center of the packet at $R = 20\,fm$, and the parameters of the Morse potential are $\alpha = 0.3\,fm^{-1}$, $r_e = 4\,fm$ and $A = 4\,fm^{-2}$. The two former are the same as used throughout this paper, but the latter is such that for this potential there is only one resonance and one bound state. The resonance is located between $1.5071 < k < 1.5072$, and the resonance momentum and width are

$$k_r \simeq 1.50716\,fm^{-1},\,and\,\Gamma \simeq 1.9 \times 10^{-5}\,fm^{-2}. \tag{31}$$

The corresponding value of the time delay is $\tau_D \simeq 2.1 \times 10^5\,fm^2$, and the unit of time is $\tau = 2.63 \times 10^4\,fm^2$. The integral in Eq. (29) for the resonance packet $\Psi_R(r,t)$ was obtained numerically in the interval $[1.5071, 1.7072]$ by using Bode's rule with equispaced momentum steps $\Delta k = 10^{-6}\,fm^{-1}$, and the result is displayed in Fig. 9. As expected, the resonance wave packet has a much smaller magnitude than the full wave packet, and its time motion is determined by the time scale $\tau$, Eq. (30a), which is more than four orders of magnitude longer than the scale $T$ of the full packet. As can be seen from the figure, the radial dependence of the resonance wave packet varies only little with time, in contrast to the full packet. This is due to the fact that in the potential valley region the contributions to the integral in Eq. (29) are dominated by the wave functions $\Phi(k,r)$ at the peak of the resonance, which differ little from each other in that region for such a narrow resonance, as is shown in Fig. 10. As a result, the resonance wave packet at $t = 0$, illustrated in Fig. 11, has a radial shape which is very similar to a second bound state wave function (if it existed) in that it has one node, and has a small magnitude beyond the barrier.

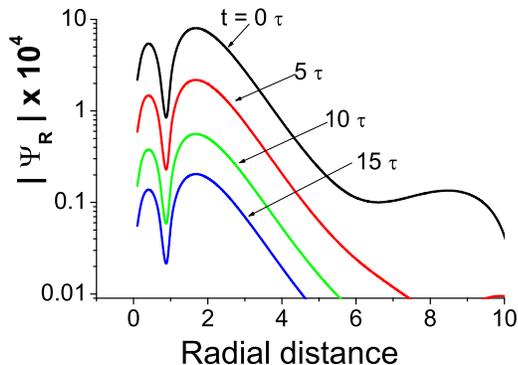

Figure 9: The absolute value of the resonance part of the wave packet defined in Eq. (29). The momentum width used to evaluate the integral is $\Delta k = 10^{-6}\ fm^{-1}$, the time scale $\tau$ is given in Eq. (30a). It is $\simeq 2.6 \times 10^4$ times longer than time scale $T$ for the full wave packet. In this log-plot the curves for the various times turn out to be spaced by approximately equal distances, showing that the magnitude of the wave packet decreases exponentially with $\tau$-time, as expected.

In summary, the resonant part of the wave packet behaves rather differently from the full wave packet. Its main probability distribution is located inside of the resonance valley at $t = 0$, and the magnitude of this distribution decays with time very slowly according to the time scale determined by the width of the resonance. Furthermore, the radial shape of the resonant portion of the wave packet resembles that of a bound state at $t = 0$.

## 5   Summary and Conclusions

This article is based on the work by two Research Experience for Undergraduate (REU) students, performed during two successive summers. The first stage consisted in showing that the known nature of the analytic solutions of the Schrödinger equation for a Morse potential could be utilized to construct a solution which vanishes at a particular point, the zero of the radial coordinate. This solution, based on hypergeometric functions with complicated complex parameters, was shown to satisfy the correct reality conditions, both for the wave function and the phase shift. It was shown independently that this solution agrees with a non-analytical numerical solution of the Schrödinger equation, and serves to provide a severe test for the accuracy of the latter. Our solution further exhibited interesting resonance properties, which could be numerically linked to the properties of the scattering $S-matrix$ in complex momentum space. In particular, the good quality of the Breit-Wigner approximation to the resonance could be confirmed, with the position and width determined entirely from the position of the poles of the $S-matrix$. Further, a simple method was developed for finding the location in momentum space of resonances, which consists in solving a two-point boundary condition eigenvalue equation in configuration space. That calculation does not require the knowledge of the wave function or of the phase shifts, but only requires the knowledge of the position in coordinate space of the barrier, where the wave function is required to vanish. It easily finds resonances, no matter how narrow their width, contrary to what is the case for other procedures.

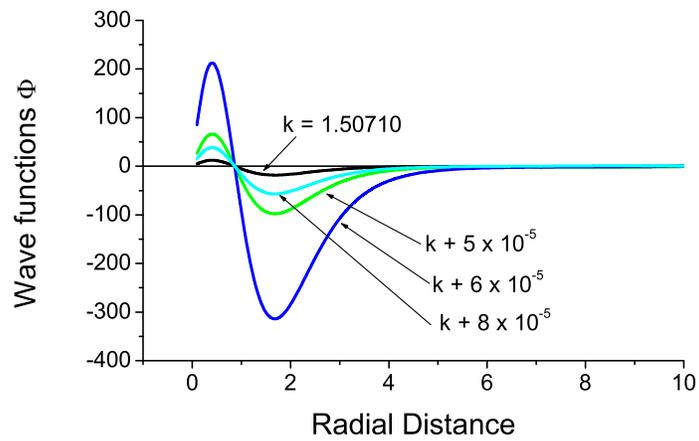

Figure 10: Radial wave functions for various values of the momenta $k$ in the resonance region of the $A = 4$ Morse potential. The numbers in the Legend are the momentum values which have to be added to $k = 1.5071$. For example, the solid curve, labeled $k + 6 \times 10^{-5}$ corresponds to $k = 1.50716\,fm^{-1}$, and is near the peak of the resonance. The curves to either side of this value of $k$ have smaller amplitudes.

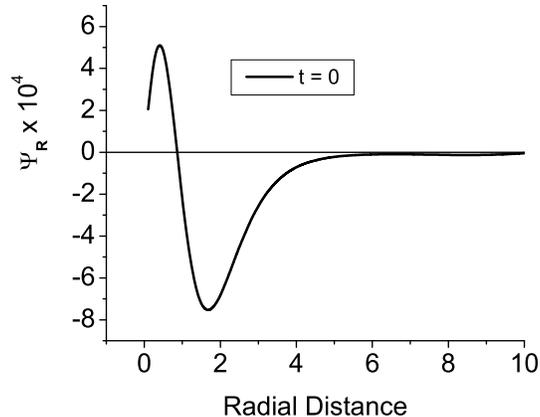

Figure 11: The resonant part of the wave packet wave function at $t = 0$. Its absolute value is also illustrated by the top line in Fig. 9. The radial shape of this function is similar to that of a second bound state, which is to be expected, since this Morse potential ($A = 4$) has only one true bound state

The numerical evaluation of the analytical expressions was done both by utilizing MATHEMATICA [15], as well as by a FORTRAN program, one serving to check the other. It thus provided an incentive for students to simultaneously learn to appreciate the power of MATHEMATICA, and the power of analytic solutions. At the same time it allowed the students to become acquainted with scattering theory by a "hands-on" procedure. It is thus hoped that this article will provide an interesting topic for courses which teach the complicated art of scientific computing [22].

# Appendix 1

The analytic results described in section 2 will be derived here. Upon transforming the variable $r$ into $z$ [1], [7] according to Eq. (8), and defining $\Phi(z) = z^{1/2} R(r)$, Eq.(3) yields for $\Phi$ the equation

$$\frac{d^2\Phi}{dz^2} + \left( -\frac{1}{4} + \frac{(k/\alpha)^2 + 1/4}{z^2} + \frac{i\beta}{z} \right) \Phi = 0, \tag{32}$$

which is of the form of a Whittacker equation, as given by Eq. (13.1.13) of Ref. [12]. Two independent solutions $\Phi$ of Eq. (32) are given by [12], which then provide two solutions of Eq. (3) of the form

$$\psi^{(\pm)}(r) = e^{-z/2} z^{\pm ik/\alpha} M(a^{(\pm)}, b^{(\pm)}; z), \tag{33}$$

where

$$a(\pm k) = a^{(\pm)} = \frac{1}{2} \pm i\frac{k}{\alpha} - i\beta \tag{34}$$

$$b(\pm k) = b^{(\pm)} = 1 \pm 2i\frac{k}{\alpha}. \tag{35}$$

It can be noted, in passing, that Eq.(32) can be transformed into an equation with real terms only, which is very similar to that of a radial Coulomb function for an energy $\Re/4$, where $\Re$ is Rydberg's constant, an effective charge $Z = \beta/2$, and a complex angular momentum $\ell^{(\pm)} = -1/2 \pm i\, k/\alpha$.

In view of Eq. (8), one has

$$\left( \frac{z}{z_0} \right)^{\pm i k/\alpha} = e^{\mp ikr}, \tag{36}$$

and in the limit $r \to \infty$, $M^{(\pm)} \to 1$ because $z \to 0$. As a result one finds that in the limit $r \to \infty$

$$(z_0)^{\mp ik/\alpha}\, \psi^{(\pm)}(r) \to e^{\mp ikr}. \tag{37}$$

which is valid whether $k$ is real or complex, and which leads to the correct sinusoidal behavior at large distances for real values of $k$.

The general solution of Eq. (3) is

$$R(r) = C^{(+)} \psi^{(+)}(r) + C^{(-)} \psi^{(-)}(r), \tag{38}$$

where the constants $C^{(\pm)}$ are determined as follows. Their ratio is determined by the requirement that $R$ vanishes at $r = 0$

$$\frac{C^{(+)}}{C^{(-)}} = -\frac{\psi^{(-)}(0)}{\psi^{(+)}(0)} = -(z_0)^{-2i\,k/\alpha} \frac{M_0^{(-)}}{M_0^{(+)}}, \tag{39}$$

where $M_0^{(\pm)} = M(a^{(\pm)}, b^{(\pm)}; z_0)$. Making use of the above expression, together with Eq. (36) one obtains

$$R(r) = e^{-z/2}\, C^{(+)}\, z_0^{ik/\alpha} \left[ M^{(+)} e^{-ikr} - \left( M_0^{(+)}/M_0^{(-)} \right) M^{(-)} e^{+ikr} \right] \tag{40}$$

In the asymptotic limit of $r \to \infty$, the S-matrix is the negative of the ratio of the coefficients of the outgoing wave $\exp(ikr)$ to that of the ingoing wave $\exp(-ikr)$. Hence, in view of 37, one obtains the result of Eq. (13). In the above the phase-shift $\delta$ is in general complex, and becomes real when the imaginary part of $k$ is zero.

The proof of properties (14) and (15) for the $S-matrix$ above is based on Eq. (13.1.27) of Ref. ([12])

$$e^{-z/2} M(a,b;z) = e^{z/2} M(b-a,b;-z). \tag{41}$$

together with the facts that in our application $z$ is purely imaginary, and the quantities $a^{(\pm)}$ and $b^{(\pm)}$ have the special properties given by Eq. (34). In view of the above, the function $\mathfrak{M}$, Eq. (10), according to Eq. (41), has the property

$$e^{-z/2} M^{(-)}(z) = [\mathfrak{M}(k^*, z)]^*, \tag{42}$$

from which it follows that Eq. (13) can also be written in the form

$$S = \frac{\mathfrak{M}(k, z_0)}{[\mathfrak{M}(k^*, z_0)]^*}. \tag{43}$$

For real values of $k$ the above equation shows that the absolute value of $S$ is unity, and that Eq. (12) is valid.

The proof of Eq. (11) also follows from Eq. (42). The normalization requirement, Eq. (7), can be implemented by requiring

$$C^{(+)} z_0^{ik/\alpha} = -e^{-i\delta}/2i$$

which, in view of Eq. (13), leads to the result

$$R(r) = -(e^{-z/2}/2i) \left[ -M^{(+)} \left( \frac{M_0^{(-)}}{M_0^{(+)}} \right)^{1/2} e^{-ikr} + M^{(-)} \left( \frac{M_0^{(+)}}{M_0^{(-)}} \right)^{1/2} e^{+ikr} \right]. \tag{44}$$

This function vanishes at $r = 0$, is in accordance with Eq. (7) for $r \to \infty$, and gives rise to Eq. (11) for real values of $k$.

# Appendix 2

In how large a region of the complex momentum plane around the pole is the Breit-Wigner approximation valid ?. By looking at Eq. (20) one is tempted to investigate this question by examining the behavior of $(S-1) \times (E_P - E)$, and expecting the answer to be equal to an $E$-independent quantity $i\Gamma$. However, because of the presence of the unknown background phase $\xi$, this procedure has to be modified by rewriting Eq. (20) in the form

$$S - 1 = e^{2i\xi}\left(e^{2i\Delta} - 1\right) + (e^{2i\xi} - 1),$$

and multiplying both sides of the equation by $(E_P - E)$, or, $k_P^2 - k^2$. In the region of validity of the B-W approximation the result should equal

$$(S - 1) \times (k_P^2 - k^2) \simeq e^{2i\xi} i\Gamma + (e^{2i\xi} - 1) \times (k_P^2 - k^2), \qquad (45)$$

in view of Eq. (20). In the above both $k_P$ and $k$ are complex quantities, with $k_P$ denoting the known location of the pole. Numerical tests were performed by varying $k$ along a circle centered at $k_P$ with a radius $\rho$, and plotting the left hand side of Eq. (45) as a function of the angle $\phi$ of the $\rho$ vector along the circle. By ignoring terms of the order of $\rho^2$ in comparison with $k_P^2$, then according to the right hand side of Eq. (45) the result should be a sinusoidal function of $\phi$, oscillating around a complex constant $e^{2i\xi} i\Gamma$, with an amplitude proportional to the radius $\rho$. Such behavior was indeed found to be the case, out to radii as large as the $2k_I^{(P)}$ This is twice the distance of the pole to the real $k-$ axis, and encompasses the location where the $S-$matrix has a zero.